\documentstyle[12pt,epsf]{article}
\begin{document}
{\small \bf
\noindent
Preprint\\
FZR--149\\
September 1996}

\vspace*{3cm}

\begin{center}
{\bf Evidence for low freeze-out temperature and large transverse flow\\
in central collisions of Pb + Pb at 158 AGeV}\\[9mm]
{\sc B. K\"ampfer}\\[6mm]
Institut f\"ur Theoretische Physik, TU Dresden, 01062 Dresden,\\
and\\
Forschungszentrum Rossendorf, PF 510119, 01314 Dresden\\[1cm]
\end{center}

\vspace*{1cm}

Utilizing a hydrodynamical model for the freeze-out stage in
heavy-ion reactions, we extract from resolved transverse hadron spectra
at midrapidity
an inverse slope parameter (temperature) $T_{Pb} =$ 120 MeV and an
averaged transverse flow velocity $v_{\perp Pb}^{aver} =$ 0.43 c
in central collisions of Pb + Pb at 158 AGeV.

\newpage

Recently the (yet preliminary) transverse momentum spectra of
$\pi^\pm$, $\pi^0$, $K^\pm$, $K_s^0$, $p^\pm$, $\Lambda$, $\bar \Lambda$
at midrapidity,
resulting in central collisions of Pb + Pb at beam energy 158 AGeV, have
been reported \cite{NA49,NuXu,WA80}.
To analyze these spectra one can utilize a hydrodynamical description
of the freeze-out stage. The concept of hydrodynamics for
describing heavy-ion collisions has a long history, and quite sophisticated
models have been proposed for modeling the hadron dynamics at freeze-out
(for a survey cf. \cite{Uli}).

The general interest in hydrodynamics is motivated by the
fact that this approach is intimately related to the use of the equation
of state of strongly interacting matter. Within such a framework
the chiral symmetry restoration or the
phase transition from hadron matter to a quark-gluon plasma appear as
particularly interesting. It is still one of the ultimate goals in the
realm of heavy-ion physics at relativistic energies to search for signals
of deconfined matter. Now, it seems that only a combination of various observables
can help to pin down information on the state of maximum density and excitation
energy. Sometimes penetrating probes, like dileptons and photons, are considered
as favorable signals for the diagnostic of the early stage in the course
of colliding nuclei. Hadrons, otherwise, carry mainly information on the
late stages of disassembling matter due to the strong interaction.
However, if initially a state of hot and dense matter is formed, then the resulting
strong pressure let the system rapidly explode, and a collective flow
develops as a consequence.

Guided by such ideas one can try to elucidate whether experimental
hadron spectra
support the predictions of thermodynamical and hydrodynamical concepts.
Indeed, in refs.~\cite{Johanna,Stachel1} it is claimed that in
silicon and gold induced reactions at BNL-AGS energies the
hadron abundances and the spectra point to thermal and
hadrochemical equilibrium. Also in sulfur induced reactions at CERN-SPS
energies this concept is anticipated
with a freeze-out temperature of $T_S =$ 160 MeV \cite{Stachel2},
which is in the region where the deconfinement transition is expected.

Here we would like to point out that the available data set
\cite{NA49,NuXu,WA80,Johanna}
in central Pb + Pb reactions at CERN-SPS energy
favors a freeze-out temperature of 120 MeV and an averaged transverse flow
velocity of 0.43 c. Therefore, the puzzling situation
arises that in the sulfur induced reactions
a much higher freeze-out temperature appears than in lead reactions.

We base our analysis on the hydrodynamical model
with linear velocity profile at freeze-out time $\tau_{f.o.}$ \cite{Uli}.
The transverse momentum distributions then read in Boltzmann approximation
\begin{equation}
\frac{d N^i}{m_\perp \, d m_\perp \, dy} = {\cal N}_i
\int_0^1 d \xi \, \xi \, m_\perp
I_0 \left( \frac{p_\perp \mbox{sh} (\rho)}{T} \right) \,
K_1 \left( \frac{m_\perp \mbox{ch} (\rho)}{T} \right),
\end{equation}
where $\rho = \mbox{arcth} (v_\perp (\xi))$,
$v_\perp(\xi) = \frac 32 v_\perp^{aver} \, \xi$, and $v_\perp^{aver}$
is the averaged transverse flow velocity;
$m_\perp = \sqrt{m_i^2 + p_\perp^2}$ denotes the transverse mass of
the hadron species $i$;
$I_0$ and $K_1$ are Bessel functions.
The normalization constants
${\cal N}_i = g_i R_{f.o.}^2 \tau_{f.o.} \lambda_i \exp\{\frac{\mu_i}{T}\}
/\pi (\hbar c)^3$
depend on the chemical potentials $\mu_i$, phase space occupation factors
$\lambda_i$ and particle degeneracies $g_i$.
This model relies on boost-invariant scaling hydrodynamics and a unique
freeze-out time in both longitudinal and transverse directions.
While the net proton rapidity density looks in Pb + Pb collisions quite
flat in the interval $1 < y < 5$, the negative hadrons
show a pronounced bell-shaped rapidity distribution \cite{NA49}.
Therefore the applicability of the model (1) is restricted
to a sufficiently narrow interval at midrapidity.
(The rapidity distributions could be adjusted by a suitable
dependence $\mu (y)$ \cite{UliS95} together with the selection of
a finite rapidity range.)

With the model (1) one can easily describe the slopes of the resolved hadron
spectra of pions, kaons, protons and lambdas \cite{NA49} by a set of parameters
$T$, $v_\perp^{aver}$, where $T(v_\perp^{aver})$ can vary over a wide
range for given slopes.
In fitting the transverse spectra, which are parametrized in
ref.~\cite{NA49} by
$\frac{d N^i}{m_\perp \, d m_\perp \, dy} = {\cal C}_i
\exp(- m_\perp/T_i)$, one observes that, for the values $T_i$ reported
in table 1 in ref.~\cite{NA49} for Pb + Pb at 158 AGeV,
all of the quoted hadron spectra measured by the NA49 collaboration
are uniquely described by $T_{Pb} =$ 120 MeV and
$v_{\perp Pb}^{aver} =$ 0.43 c.
This situation is different from S + S and S + W, Au, where such a focus
of the curves $T(v_\perp)$ for the various hadrons does not occur
\cite{Uli}. In this respect the Pb + Pb data point to a clear flow
signal for the first time.
Our fitted spectra together with the data from
ref.~\cite{NA49} are displayed in fig.~1
(for details of acceptance corrections, error bars and rapidity binning cf.
\cite{NA49}).
Most remarkably is the change of the spectra's shapes when changing the
hadron masses.

As cross check one can compare with the preliminary
$\pi^0$ spectrum of the WA98
collaboration \cite{WA80} and finds very good agreement in the available transverse
momentum range of $p_\perp =$ 0.5 - 2.5 GeV. We emphasize that our fit
describes very well the data shown in ref.~\cite{Johanna},
namely all negatively charged hadrons for
$m_\perp - m_\pi =$ 0.015 - 1.85 GeV, $K^+$ for
$m_\perp - m_K =$ 0.015 - 1 GeV, and the net baryons $(+) - (-)$ for
$m_\perp - m_p =$ 0.015 - 1.42 GeV.
In addition, the preliminary NA44 data \cite{NuXu}
of $p^\pm$ and $K^\pm$
in the ranges $m_\perp - m_p =$ 0.01 - 0.87 GeV and
$m_\perp - m_K =$ 0.01 - 1.12 GeV, respectively,
strongly support our values of $T_{Pb}$ and $v_{\perp Pb}^{aver}$.
The situation is quite different for the preliminary NA44 $\pi^\pm$ data
\cite{NuXu}. In the low $m_\perp$ range, which is not covered by the
NA49 data, there is a pronounced leveling off not described by our model.
A careful analysis of feeding by resonance decays
is required to decide whether a large chemical potential of pions
\cite{Kataja} is needed for describing these details.
First estimates of the influence of resonance decays \cite{Slotta}
point to a possible shift of the freeze-out temperature to slightly
larger values, but leave the conclusion on flow unaltered.

A possible interpretation of the values of $T_{Pb}$ and $v_{\perp Pb}^{aver}$,
with respect to the difference to the values
$T_S$ and $v_{\perp S}^{aver} =$ 0.27 c
extracted for sulfur induced reactions at CERN-SPS energies \cite{Stachel1},
could be that the larger system Pb + Pb develops more collectivity and stays
for a longer time in contact up to freeze-out temperature, while
the smaller systems S + W, Au disassemble earlier at lower transverse
flow velocity but higher mean kinetic energy (temperature) of the hadrons.
Interestingly, our dynamical code \cite{BK}, which employs a linear transverse
velocity profile and a resonance gas model equation of state and initial
conditions appropriate for lead reactions, consistently results
in $v_\perp^{aver} =$ 0.45 c.

Here we do not speculate on a possible chemical equilibrium. In this respect
the normalization factors ${\cal N}_i$ need still an interpretation by studying
the chemical freeze-out, which does not need to be identical with thermal
freeze-out \cite{Uli}.
By the above definition of ${\cal N}_i$
one can try to deduce some combinations of chemical potentials
and phase space occupation factors. However, at the present early stage of the
data analyses one could come to less reliable results.
For instance, the rapidity densities
$d N_i /dy$ obtained from a $m_\perp$ integration of our fitted spectra
and experimental data fits in ref.~\cite{NA49} differ up to a factor 2.
Nevertheless in a first attempt \cite{Johanna} to interpret the hadron
yield ratios, a chemical freeze-out temperature of 160 MeV is found.
While this value, when identified with thermal freeze-out,
is consistent with the transverse spectra of all negative hadrons,
$K^+$, and net baryons for the model (1) with
$v_\perp^{aver} =$ 0.3 c \cite{Johanna}, it is disfavored essentially by the
proton and lambda data \cite{NA49,NuXu}.

In ref.~\cite{Mueller} the dragging coefficient between pions and nucleons
is calculated and is found to be too small that pions and nucleons are
likely to flow with the same velocity. In this respect the circumstantial
evidence for a common transverse flow of all hadrons could be traced back
to emerge from an early overdense (and possibly deconfined) stage, which
triggers the onset of the flow. But it should be emphasized that
more accurate data, in particular in a wider transverse momentum range,
are needed to come to firm conclusions. A finer rapidity binning
is desirable to understand the longitudinal dynamics, too.
Also feeding by resonance decays must be included in an advanced analysis.

In summary, within a hydrodynamical model
we extract from the transverse momentum spectra of resolved
hadrons a thermal freeze-out temperature $T_{Pb} =$ 120 MeV and averaged transverse
flow velocity $v_{\perp Pb}^{aver} =$ 0.43 c in central Pb + Pb collisions
at 158 AGeV.

{\bf Acknowledgments:}
M. Ga\'zdzicki and N. Xu are gratefully acknowledged for informing me
on the data analyses of the NA49 and NA44 collaborations prior to publication.
I am indebted to J. Cleymans, U. Heinz, B. M\"uller, O.P. Pavlenko,
K. Redlich, and J. Stachel
for useful conversations on transverse flow, and
to C. Slotta for informing me on his estimates of the influence of
resonance decays.
This work is supported by BMBF grant 06DR666.

\begin{center}
{\bf Figure caption}
\end{center}
Fig.~1:
Fits of the resolved transverse momentum spectra of hadrons
(a: $\pi^\pm$, b: $K^\pm$, c: $K^0_s$, d: $p^\pm$, e: $\Lambda$,
$\bar \Lambda$).
The preliminary data are from ref.~\cite{NA49}.
Our normalization factors ${\cal N}_i$ are given in the keys.
\begin{figure}[t]
  \epsfxsize=6cm
  \epsffile{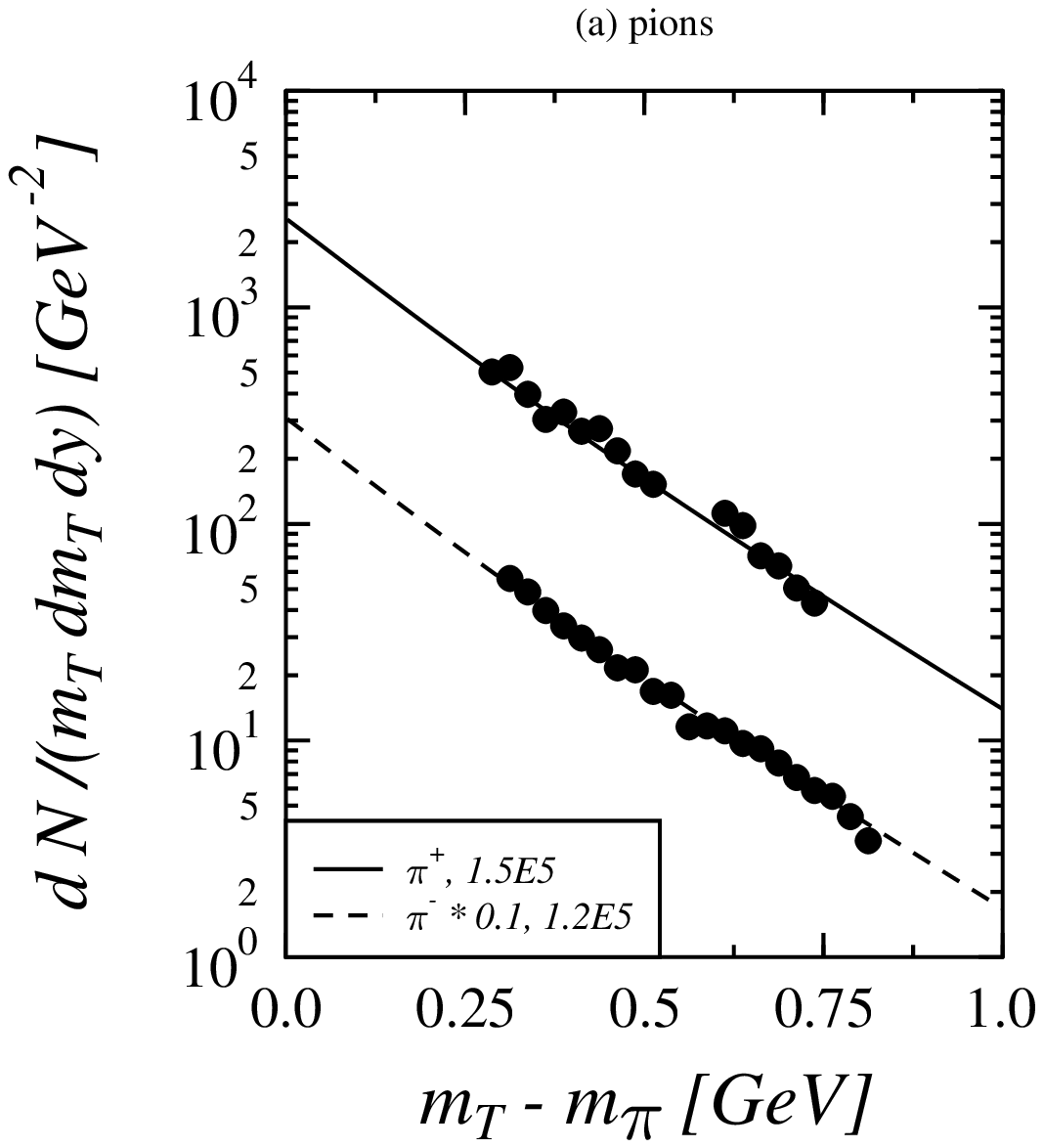}
\end{figure}
\begin{figure}[t]
  \epsfxsize=6cm
  \epsffile{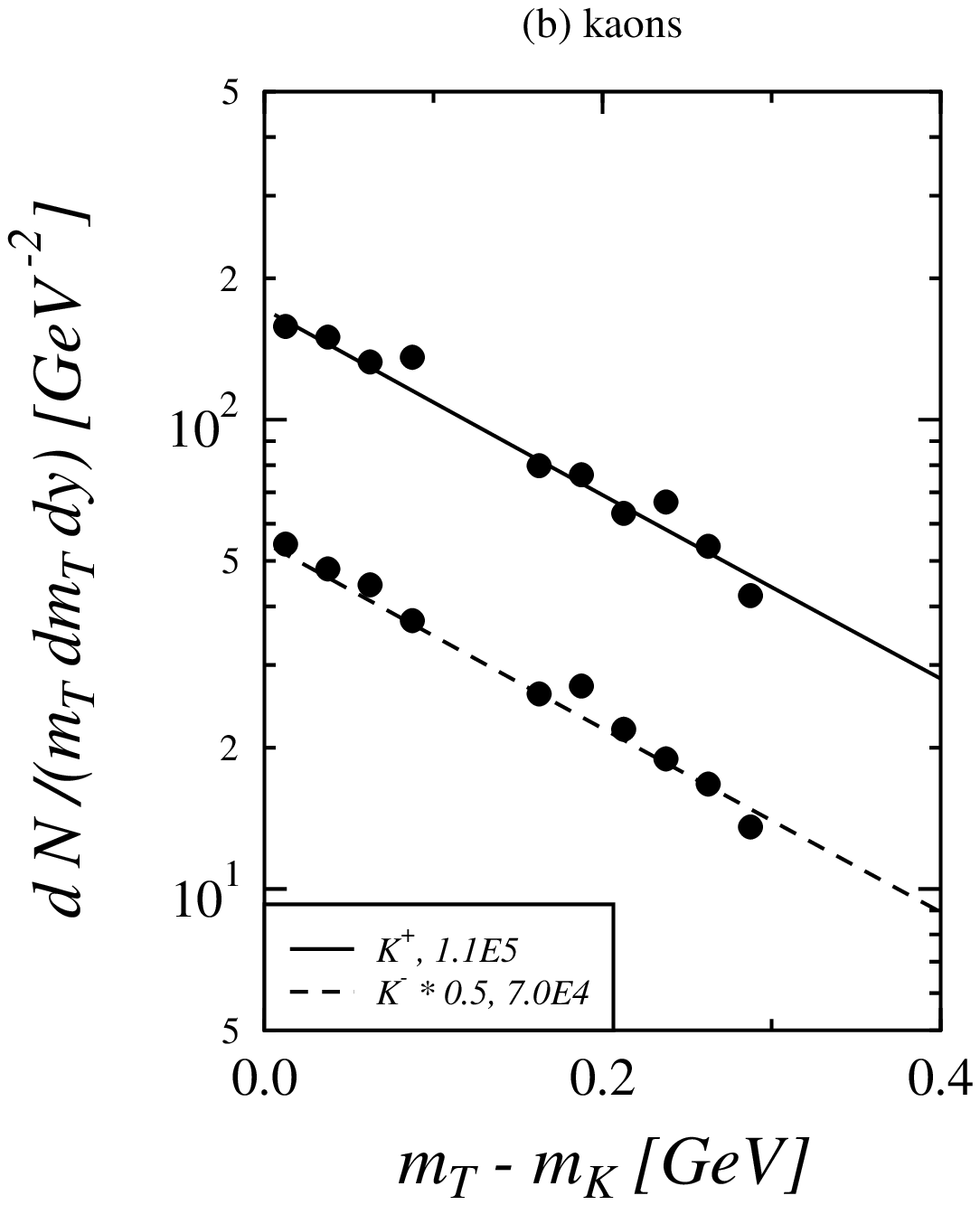}
\end{figure}
\begin{figure}[t]
  \epsfxsize=6cm
  \epsffile{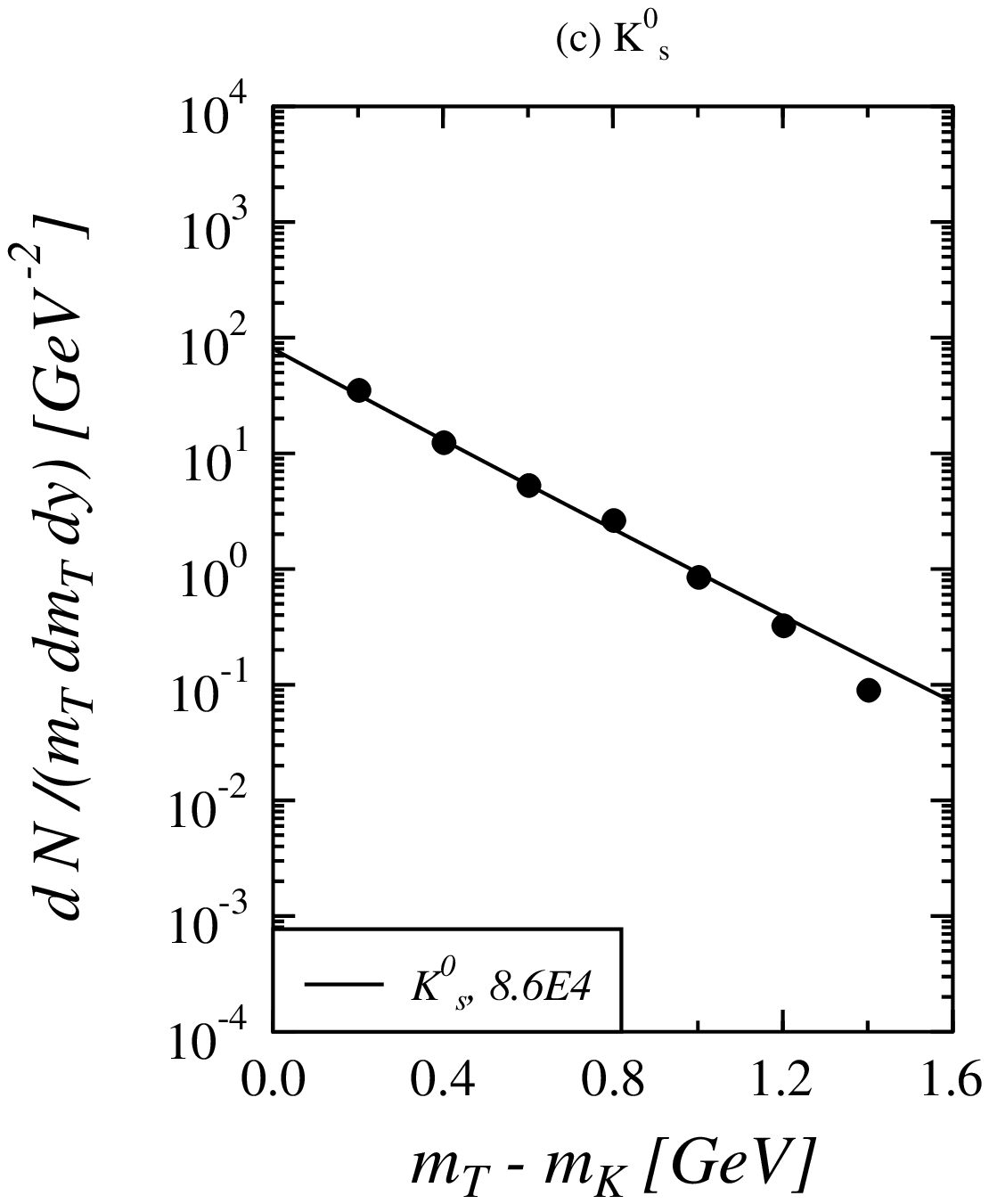}
\end{figure}
\begin{figure}[t]
  \epsfxsize=6cm
  \epsffile{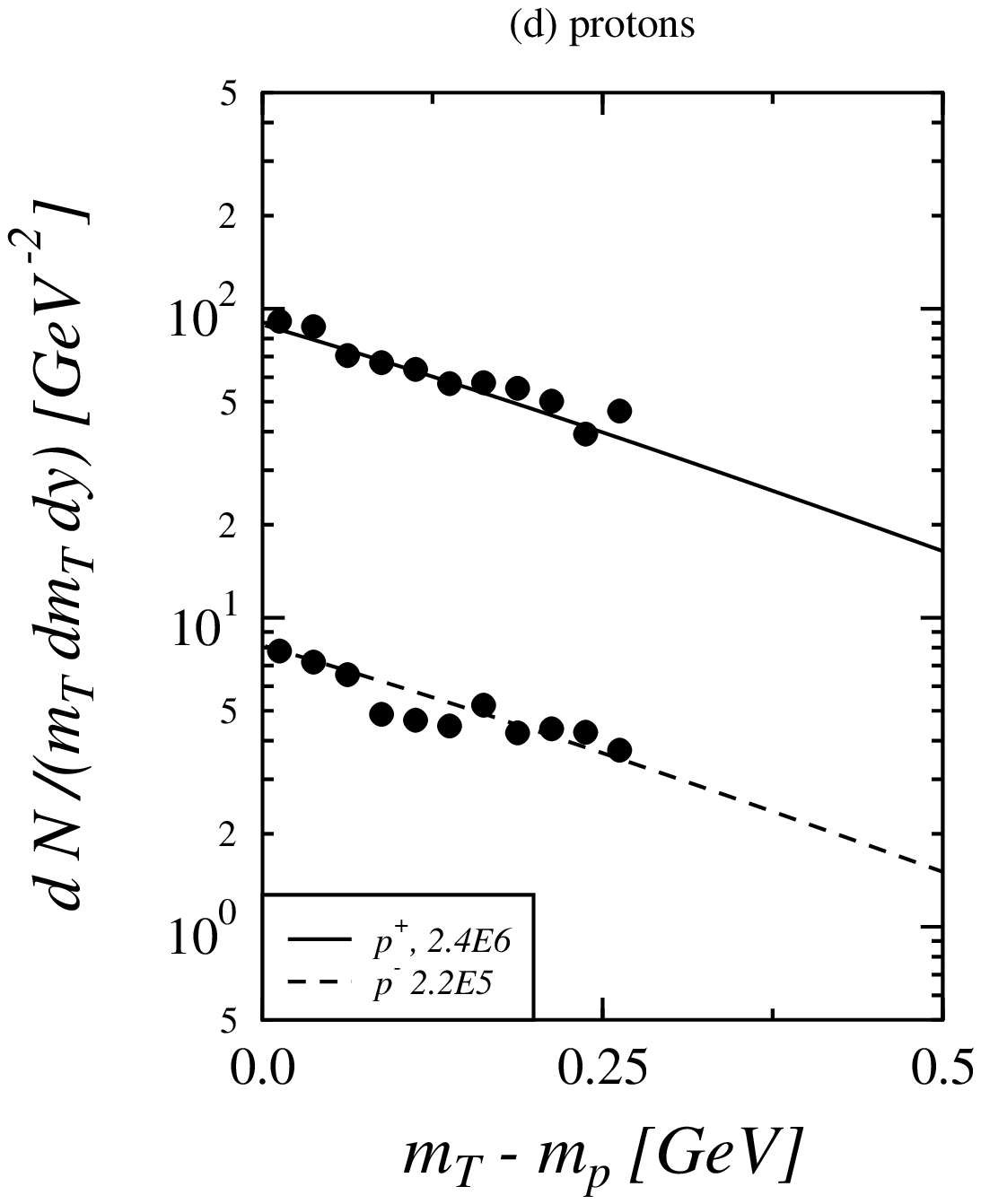}
\end{figure}
\begin{figure}[t]
  \epsfxsize=6cm
  \epsffile{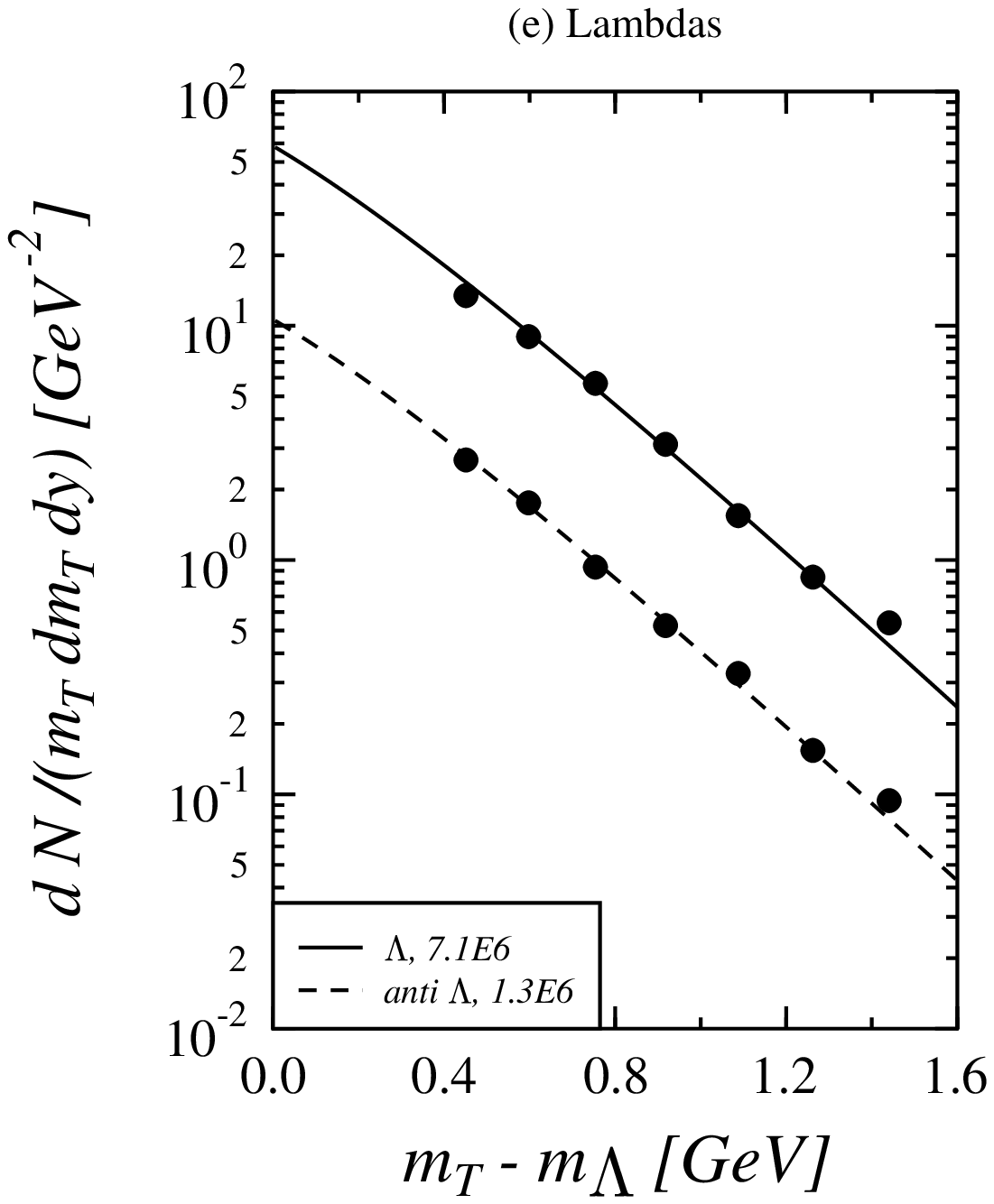}
\end{figure}
\end{document}